\documentclass[12pt,preprint]{aastex}

\renewcommand {\deg}   {\mbox{$^\circ$}}

\newcommand   {\kms}   {\mbox{km\,s$^{-1}$}}
\renewcommand {\ga}    {\mbox{\rlap{\hbox{\lower5pt\hbox{$\sim$}}}\hbox{$>$}}}
\renewcommand {\la}    {\mbox{\rlap{\hbox{\lower5pt\hbox{$\sim$}}}\hbox{$<$}}}

\newcommand  {\solar}  {\mbox{$_{\odot}$}}

\begin{document}



\def\kms {\hbox{km{\hskip0.1em}s$^{-1}$}} 
\def\msol{\hbox{$\hbox{M}_\odot$}}
\def\lsol{\hbox{$\hbox{L}_\odot$}}
\def\kms{km s$^{-1}$}
\def\Blos{B$_{\rm los}$}
\def\etal   {{\it et al. }}                     
\def\psec           {$.\negthinspace^{s}$}
\def\pasec          {$.\negthinspace^{\prime\prime}$}
\def\pdeg           {$.\kern-.25em ^{^\circ}$}
\def\degree{\ifmmode{^\circ} \else{$^\circ$}\fi}
\def\ee #1 {\times 10^{#1}}          
\def\ut #1 #2 { \, \textrm{#1}^{#2}} 
\def\u #1 { \, \textrm{#1}}          
\def\nH {n_\mathrm{H}}

\def\ddeg   {\hbox{$.\!\!^\circ$}}              
\def\deg    {$^{\circ}$}                        
\def\le     {$\leq$}                            
\def\sec    {$^{\rm s}$}                        
\def\msol   {\hbox{$M_\odot$}}                  
\def\i      {\hbox{\it I}}                      
\def\v      {\hbox{\it V}}                      
\def\dasec  {\hbox{$.\!\!^{\prime\prime}$}}     
\def\asec   {$^{\prime\prime}$}                 
\def\dasec  {\hbox{$.\!\!^{\prime\prime}$}}     
\def\dsec   {\hbox{$.\!\!^{\rm s}$}}            
\def\min    {$^{\rm m}$}                        
\def\hour   {$^{\rm h}$}                        
\def\amin   {$^{\prime}$}                       
\def\lsol{\, \hbox{$\hbox{L}_\odot$}}
\def\sec    {$^{\rm s}$}                        
\def\etal   {{\it et al. }}                     

\def\xbar   {\hbox{$\overline{\rm x}$}}         

\shorttitle{}
\shortauthors{}

\title{ALMA Observations of the Galactic Center:\\
SiO  Outflows and High Mass Star Formation near Sgr A*}
\author{F. Yusef-Zadeh$^1$, M. Royster$^1$,  M. Wardle$^2$, R. Arendt$^3$, H. Bushouse$^4$,
D. C. Lis$^5$, M. W. Pound$^6$, D. A. Roberts$^1$, B. Whitney$^7$ and   A. Wootten$^8$
}
\affil{$^1$Department of Physics and Astronomy and Center for Interdisciplinary Research in Astronomy,
Northwestern University, Evanston, IL 60208}
\affil{$^2$Department of Physics and Astronomy, and Research Centre for Astronomy,
Astrophysics \& Astrophotonics, Macquarie University, Sydney NSW 2109, Australia}
\affil{$^3$CREST/UMBC/NASA GSFC, Code 665, Greenbelt, MD 20771}
\affil{$^4$Space Telescope Science Institute, 3700 San Martin Drive, Baltimore, MD  21218}
\affil{$^5$California Institute of Technology, MC 320-47, Pasadena, CA 91125}
\affil{$^6$University of Maryland, Department of Astronomy,  MD  20742}
\affil{$^7$Space Science Institute, 4750 Walnut Street, Suite 205, Boulder, CO 80301}
\affil{$^8$National Radio Astronomy Observatory,  Charlottesville, VA 22903}


\begin{abstract} 
ALMA observations of the Galactic center with  spatial resolution  $2.61''\times0.97''$ resulted in the 
detection of  11  SiO (5-4) clumps of molecular gas  
within 0.6pc (15$''$) of Sgr A*,  
interior to the 2-pc circumnuclear molecular ring.
The three SiO (5-4) clumps closest to Sgr A* 
show  the largest central velocities,  $\sim150$ \kms,  and broadest asymmetric linewidths 
with full width zero intensity (FWZI) $\sim110-147$ \kms. The remaining clumps, 
 distributed mainly  to the NE of the ionized 
mini-spiral,  have narrow FWZI  ($\sim18-56$ \kms). Using CARMA SiO (2-1) data, 
LVG modeling of the 
the SiO  line ratios 
for  the broad velocity clumps,
constrains the column density N(SiO) $\sim10^{14}$ cm$^{-2}$, 
and the H$_2$ gas density n$_{\rm 
H_2}=(3-9)\times10^5$ cm$^{-3}$ for an assumed kinetic temperature 100-200K. 
The SiO clumps are interpreted as highly embedded 
protostellar outflows, signifying an early stage of massive star formation 
near Sgr A* in the last 
$10^4-10^5$ years. 
Support for this interpretation 
is provided  by the  SiO (5-4)  line luminosities and velocity widths which lie in the range 
measured for  protostellar outflows in star forming regions in the Galaxy. 
Furthermore, SED modeling of stellar sources shows two YSO candidates near SiO clumps, 
supporting in-situ star formation near 
Sgr A*. 
We discuss the nature of star formation  where the gravitational potential of the black hole dominates. 
In  particular, we suggest that external  radiative pressure exerted on self-shielded molecular clouds enhances the gas 
density, before the gas cloud   become gravitationally unstable near Sgr A*.
Alternatively, collisions between clumps in the ring may 
trigger gravitational collapse. 
\end{abstract}


\keywords{Galaxy: center - clouds - ISM: general - ISM - radio continuum - stars: protostars}

\section{Introduction}

The compact radio source Sgr A*  located at the very 
dynamical center of our galaxy coincides with a $4 \times 10^6$\msol\, black hole (Ghez 
\etal  2008; Gillessen \etal  2009). 
Star  formation near Sgr A* is forbidden, unless the gas 
density is large enough  for self-gravity 
to overcome    the strong tidal shear of the back hole.   
One viable mechanism is cloud capture to  form  a massive 
and dense gaseous disk around the black hole. 
The disk becomes  gravitationally unstable, 
creating  a new generation of stars (e.g., Levin and Beloborodov 2003; Nayakshin \etal 2007;
Paumard \etal 2006; Lu \etal 2009; Mapelli \etal 2012; Wardle \& Yusef-Zadeh 2008; 
Bonnell \& Rice 2008; Alig \etal 2012).  
This formation scenario has been applied to 
disk-like distribution of 
young stars on a scale of 0.03-0.3 pc, to the 
2-5pc circumnuclear molecular ring  (CND or CNR) orbiting Sgr A* as well as to 
AGNs with megamaser 
disks (Wardle \&  Yusef-Zadeh 2008; 2012). The Galactic center provides an opportunity for 
testing models of stellar birth with far reaching implications on the nature of star 
formation in the nuclei of galaxies hosting massive black holes.

Because  the age of the stellar disk  orbiting Sgr A* is several million years, it is not 
possible to identify signatures of early phases of star formation such as maser 
activity near the black hole. 
However, a 
 number of recent studies suggest infrared excess sources in the N arm,  as well as young 
stellar object (YSO) candidates in IRS 13N within a projected distance of 0.12 pc (3$''$) from Sgr A* 
(e.g., Viehmann \etal 2006; Eckart \etal 2012; 
Nishiyama and Sch\"odel 2013). 
These measurements 
imply that star formation has taken  place within $\sim10^5$ years. 
On a larger scale of $50''-100''$,  or projected distance of 2-4 pc 
(assuming that the distance to the Galactic center is 8.5 kpc),  the molecular ring 
(Jackson \etal 1993; 
Montero-Castano \etal 2009, Martin \etal  2012) that encircles Sgr A* with a rotational velocity 
of $\sim100$ \kms\, shows signatures of massive star formation activity within the last 
$\sim10^4 - 10^5$
 years (Yusef-Zadeh \etal 2008).
Here, we present the earliest signatures of on-going star formation on a  scale of about 11$''$ (0.44 pc) 
from Sgr A*: the presence of SiO (5-4) line emission around  the interior of the molecular ring. 
The SiO molecule is an {\it{excellent}} tracer of protostellar outflows and is shock excited,  since 
silicon is removed from dust grains, significantly increasing gas-phase abundance. 




\section{Observations and Data Reduction} 
{\bf {ALMA}}: 
The data were obtained through the ALMA Science Verification process on 
June 26, 2011.  
Utilizing  only twelve of the 12 m antennas, the observations consisted of a 7-point mosaic 
at the position of Sgr A*: $(\alpha, \delta)= 17^h 45^m 40^s.04, -29^{\circ} 0' 28''.12$.  
Titan was initially used as the flux calibrator and 3C279 as the 
initial bandpass and phase calibrator.  NRAO 530 was observed periodically 
to correct for any changes in phase and amplitude as 
a function of time.  
Four basebands of total width 1.875 GHz were used with the high spectral resolution, 
 which 
contained 3840 channels each.  The spectral windows were centered at roughly: 216.2 GHz, 
218.0 GHz, 231.9 GHz, and 233.7 GHz. 
The SiO (5-4) line emission is centered at 217.105 GHz.  
The editing and calibration of the data was carried 
out by the Science Verification team in CASA.  The imaging was completed by modifying the supplied imaging script.  SiO 5-4 fell at 
the intersection of the first two basebands at 216.2 GHz and 218.0 GHZ.  
As a result, care was taken to note any changes in amplitude 
and phase at the edge channels in combining the two windows.  A linear continuum was fitted and subtracted 
from the line-free channels 
before CLEANing.  
The data was averaged by a factor of five to achieve a spectral resolution of 3.371 \kms\, (2.441 MHz) using  the raw binning of 
0.674 \kms\,(0.488 MHz).  To avoid confusion with nearby lines, we studied velocity structures within 440 \kms\, 
spatial resolutions of $2.61''\times0.97''$.




{\bf {CARMA}}: 
The SiO (2-1) line data were taken with
the Combined Array for Research in Millimeter-wave Astronomy (CARMA)
during the 2009 and 2010 observing
seasons in the D and C array configurations.  The array consisted of
six 10.4m antennas and nine 6.1m antennas and the maps were made on a
127-point hexagonal mosaic, Nyquist-sampling the 10.4m antenna primary
beam.  The spatial resolution and
spectral resolutions of the final maps are $8.87\arcsec\times 4.56\arcsec$
and 6.74 \kms, respectively. 


\section{Results and Discussion}

Figure 1a shows a composite 3.6 cm continuum image of the three arms of the mini-spiral (Sgr A West) surrounded by the CNR as 
traced by HCN line emission (Christopher \etal 2005). The inner 1$'$ (2.4pc) of the CNR, as observed with ALMA, shows a large 
concentration of molecular clumps in the SiO (5-4) line emission. The distribution of molecular and ionized gas orbiting Sgr 
A* is generally described as a central cavity that consists of ionized gas of Sgr A West coupled to the surrounding 
molecular ring (e.g.  Roberts and Goss 1993; Serabyn and Lacy 1985). A string of SiO clumps is found in the region between the 
N and E arms of Sgr A West.  A close up view of these clumps is shown in Figure 1b. 
 What is most interesting is the presence of SiO clumps in the interior of 
the ring, which is expected to be dominated by the central cavity of ionized gas and atomic gas traced by [OI] emission, 
devoid of any molecular gas.  
Figure 1b also shows two clumps (green), with velocities $v_{\rm r}\sim$ 136 to 160 \kms, 
and FWHM linewidths 47 to 55 \kms. These ALMA clumps, 1 and 11, are located about $7''$ NE and 11$''$ SW of Sgr A*, 
respectively, and do not follow the kinematics of the molecular ring orbiting Sgr A*. Figure 1c shows the positions of these 
highly redshifted velocity clumps on a 3.6$\mu$m image taken with the VLT.  Clumps 1 and 2 lie $\sim2''$ west of IRS 1W, a 
mass-losing W-R star with a bow shock structure (Tanner \etal 2003).




To illustrate the kinematics of SiO (5-4) molecular clumps, 
the spectra of broadest velocity clumps  are shown in Figure 1d whereas the 
the spectra of 8 clumps with narrow linewidths within the ring are shown  
in Figure 2 where the  distribution of SiO (5-4) peak line emission is shown. 
Columns 1 to 7 of Table 1 list the source numbers, coordinates, intensity of the peak emission, Gaussian fitted 
central velocity,  FWHM linewidths,   and FWZI of the 11 sources labeled in Figure 2. 
FWZI linewidths are listed  because many  sources show velocity  
profiles with blue-shifted wings (e.g., clumps 1, 2, 11). 
Clumps  4  to 8  run parallel to the 
eastern edge of the N arm of ionized gas showing typical linewidths  between 11 and 21 \kms and peak 
radial velocities that decrease from 37.8 \kms\, at Clump 8 to 7.4 \kms at Clump 4. The trend in 
radial velocity change in molecular clumps is consistent with the trend  in the kinematics 
of ionized gas of the N arm (e.g., Zhao \etal 2011). 
However, the central velocities of ionized in the N arm 
are  $\sim$100 \kms\,   and decrease to 0 \kms\, as the ionized gas approaches closer to Sgr A* (e.g., 
Roberts and Goss 1993; Jackson \etal 1993;  
Zhao \etal 2011). 
The kinematics of SiO (5-4) 
is  dissimilar to that  of the ionized gas of the N arm 
and the "tongue" of 
neutral [OI] gas. The spatial and velocity distributions 
of  SiO (5-4) suggest a clumpy and  dense molecular cloud
lying  in  the interior of the  molecular ring.


 


Clumps 1, 2 and 3  defy the kinematical trend
of the N arm:
they display  highly red-shifted ($<$160 \kms) and blue-shifted velocity  (-28 \kms)
components and lie adjacent to the ionized  gas of the N arm, but with dissimilar velocities. 
Figure 3a shows contours of highly red-shifted SiO (5-4) line emission from Clumps 1 and 2
superimposed on a 3cm continuum image of the N arm.
The SiO line emission from Clump 1 appears to be elongated to the NE
with non-Gaussian velocity profiles having blue-shifted wings. 
Another morphological structure of the N arm is the  wavy
structure along the direction of the flow 
(Yusef-Zadeh and Wardle 1993; Zhao \etal 2012).
Figure 3b shows contours of SiO (5-4) emission from Clump 3 superimposed on a 3cm
continuum image. 
The spectrum of Clump 3 that lies adjacent to the wavy structure
shows a radial  velocity peaking at v$\sim-28$ \kms\, and linewidth of $v_{\rm FWHM}\sim27$ \kms~ 
with a broad blue-shifted wing extending to $-100$ \kms. 
This radial velocity is inconsistent with circular motion orbiting Sgr A*. 
Several radio dark clouds  (Yusef-Zadeh 2012)
lie to the east of the N arm in Figures 3a,b
where Clumps 1, 2 and 3 lie. These dark features trace molecular gas
and are embedded in the ionized gas of the N arm.
Although the morphology of thermal radio emission 
at the location of Clumps 1 and 3 
suggests that the 
flow of ionized gas is distorted as a result of its  interactions with molecular clumps,  
the kinematics of the molecular and ionized gas 
are inconsistent with the interaction picture. 





\begin{deluxetable}{rcccccc}
\label{Catalog}
\tablecaption{Gaussian Line Parameters of Fitted SiO Sources}
\tabletypesize{\scriptsize}
\tablewidth{0pt}
\tablehead{
\colhead{Source} & \colhead{RA}      & \colhead{DEC}    & \colhead{Peak}     & \colhead{Center}      & \colhead{FWHM} & 
\colhead{FWZI}\\
                 & \colhead{(J2000)} & \colhead{(J200)} & \colhead{mJy/beam} & \colhead{km s$^{-1}$} & \colhead{km s$^{-1}$} 
& \colhead{km s$^{-1}$}
}\startdata
 1 & 17:45:40.51 & -29.0.28.78 & 38.91 $\pm$ 2.28 &  147.81 $\pm$ 1.35 & 47.42 $\pm$ 3.29 & 110 $\pm$ 40\\
 2 & 17:45:40.74 & -29.0.26.11 & 13.92 $\pm$ 1.36 &  159.81 $\pm$ 2.60 & 54.81 $\pm$ 6.39 & 120 $\pm$ 41\\
 3 & 17:45:40.61 & -29.0.23.90 & 35.43 $\pm$ 1.84 &  -27.86 $\pm$ 0.69 & 27.11 $\pm$ 1.66 & 55 $\pm$ 19\\
 4 & 17:45:41.01 & -29.0.27.03 & 77.75 $\pm$ 3.17 &    7.98 $\pm$ 0.23 & 11.52 $\pm$ 0.55 & 25 $\pm$ 9\\
 5 & 17:45:41.06 & -29.0.24.09 & 62.14 $\pm$ 3.19 &   14.28 $\pm$ 0.28 & 11.15 $\pm$ 0.66 & 18 $\pm$ 9\\
 6 & 17:45:41.17 & -29.0.17.84 & 67.20 $\pm$ 3.64 &   20.85 $\pm$ 0.34 & 12.92 $\pm$ 0.81 & 29 $\pm$ 10\\
 7 & 17:45:41.03 & -29.0.14.25 & 75.74 $\pm$ 2.80 &   38.98 $\pm$ 0.26 & 14.66 $\pm$ 0.63 & 31 $\pm$ 12\\
 8 & 17:45:40.83 & -29.0.12.78 & 48.81 $\pm$ 2.80 &   39.96 $\pm$ 0.44 & 15.65 $\pm$ 1.05 & 31 $\pm$ 13\\
 9 & 17:45:40.03 & -29.0.12.87 & 62.31 $\pm$ 2.53 &   77.80 $\pm$ 0.38 & 19.44 $\pm$ 0.93 & 40 $\pm$ 17\\
10 & 17:45:40.31 & -29.0.43.77 & 71.32 $\pm$ 2.75 &  -42.98 $\pm$ 0.39 & 21.02 $\pm$ 0.95 & 56 $\pm$ 18\\
11 & 17:45:39.51 & -29.0.36.96 & 50.43 $\pm$ 1.81 &  136.16 $\pm$ 0.96 & 55.51 $\pm$ 2.43 & 147 $\pm$39
\enddata
\end{deluxetable}

{\bf {Protostellar  Outflows}}: 
To examine the physical properties of SiO clumps, we focus on 
the SiO (5-4) and (2-1) line profiles, as shown in Figure 3c, with 
peak line emission T$_{\rm MB}\sim0.95$ and 0.65 K, for Clump  1 and T$_{\rm MB}\sim1.1$ and 1 K, for Clump 11 at center 
velocity of $\sim147$ and $\sim136$ \kms, respectively. The similarity of SiO (5-4) and (2-1) line profiles toward Clumps 1 
and 11 is a classic signature of one-sided molecular outflows in star forming regions (e.g., Plambeck \& Menten 1990). 
We  used  the Large Velocity Gradient (LVG) model  to  constrain the 
density and column density.  
The result of this analysis is shown in Figure 3d.  
Assuming that the  
beam filling factor is  0.5, thus,  1 mJy is equivalent to 
T$_{\rm MB}=20.5$  and 7.96 mK for   SiO (5-4) and SiO (2-1) data, respectively, 
the gas density of 
hydrogen nuclei is constrained to 
n$_{\rm H_2}\sim (3-9)\times10^5$ cm$^{-3}$ within a temperature range 100-200K. 
The synthesized beam is highly elongated in one direction, thus we chose a beam filling factor of 0.5. 
Multi-transition CO and NH$_3$ 
studies of the inner 2pc  derive kinetic temperature 100-200K 
and warm gas in the 
interior of the molecular ring (Bradford \etal 2005; Herrnstein and Ho 2005).
N(SiO)/$\Delta v\sim1.7-2.4\times10^{12}$ cm$^{-2}$ \kms\, 
is better constrained  than the gas density, 
as shown in the bottom panel of Figure 3d. 
The column densities  of SiO for Clumps 1 and 11 are estimated to be 
N(SiO)$\sim8.6\times10^{13}$  and $\sim1.3\times10^{14}$ 
cm$^{-2}$, respectively. 
The total mass of molecular gas is estimated to be $\sim0.2$ \msol\, using clump radius 0.03pc (0.79$''$) for an
 unresolved sources and  n$_{\rm 
H_2}=10^6$ cm$^{-3}$. This corresponds to a column density of 1.8$\times10^{23}$ cm$^{-2}$ and SiO abundance of 
5.6$\times10^{-10}$ which is clearly much higher than is expected in quiescent molecular clouds. 
Using the size and the FWHM  velocity widths, the velocity gradient is estimated to be 833 and 165 \kms\, 
pc$^{-1}$ for clumps with broad ($\sim50$ \kms) and narrow ($\sim10$ \kms) linewidths respectively.  These imply that the clumps 
can not be bound by self-gravity because the collapse time scale is 12-58 times longer than the dynamical time scales t$_{\rm 
dyn}\sim (0.3-1.5)\times 10^3$ years, thus are likely to be outflows from YSOs. The  blue-shifted velocity wings 
and non-circular radial velocities in orbital motion about Sgr A* 
is also consistent with SiO outflows in which the approaching side of the outflow has burst through the 
edge of a molecular cloud.



One strong piece of evidence that the SiO sources are YSO outflows is that the luminosities and velocity widths lie in the range 
detected from protostellar outflows in star forming regions. 
 Figure 4 compares 
the SiO (5-4) luminosity and FWZI 
for the detected sources inside the molecular ring, to those  for samples of low-mass and high-mass YSOs 
(Gibb  \etal  2004 and 2007). 
The SiO 5-4 luminosities for the low-mass and high-mass YSO samples tabulated in Gibb \etal (2004, 2007)  
have been corrected by multiplying by 1500 and 170, respectively (A. Gibb, private communication).
All three samples show the same wide range of velocity widths, and the luminosities of SiO clumps interior to 
the molecular ring fall in between those of the low-mass and high-mass YSOs.  




What sources drive the  outflows? 
We identified two new YSO candidates in the N arm near Clumps 1 and 3. 
Figure 3e shows contours of SiO (5-4) emission from Clumps 1 and 5 
superimposed on a ratio map of L (3.6$\mu$m) to K (1.6 
$\mu$m) band images from VLT observations. The ratio map shows the dusty environment of the N arm.
 The crosses coincide with the 
positions of YSO candidates   with the positional uncertainty of 1.18$''$ at 8$\mu$m (Ramirez \etal 2008). 
The SEDs of these 
sources are analyzed by comparing a set of SEDs enhanced  by a large grid of YSO models (Whitney \etal 2003; Robitaille \etal 
2007), as Figure 3f shows their fitted SEDs.
The source 526311,  which is selected from their IR colors, 
is classified as a Stage I YSO candidate,  whereas the red source
526817 is a YSO candidate but its classification is uncertain. 
These YSO candidates have  typical ages of 10$^5$ years. 
By fitting the SEDs, we derive masses of $34.3\pm5.9$ and 19.4$\pm2.5$ \msol, luminosities $1.8\pm0.7\times10^5$ and 
$4.4\pm1.5\times10^4$ \lsol\, and mass-loss rates 5.1$\pm0.1\times10^{-4}$ and  
2.5$\pm0.1\times10^{-4}$ \msol\, yr$^{-1}$ for clumps 1 and 5, respectively. 
YSO candidate 526817 coincides with the brightest source 
IRS 10E (Viehmann \etal  2006) at the center of Clump 3 whereas 
the YSO candidate 526817 is an unresolved component of multiple sources
in IRS 1.  The proximity of massive YSO candidates, SiO (5-4) Clumps near IRS 1 and IRS 10 clusters 
containing W-R stars,  
suggest that there is still on-going star formation near these stellar clusters. 
Similar distribution of massive YSOs, and W-R stars are found in 
the IRS 13  cluster (Fritz \etal 2012; Eckart \etal 2012).







{\bf {Star Formation Mechanism}}: 
We have detected the presence of several SiO (5-4) clumps  within a pc of Sgr A*.  
Their SiO (5-4) luminosities, non-Gaussian velocity profiles, 
large linewidths unbound   by self-gravity and 
dense gas, as traced by dark radio clouds,  
all point to  the  conclusion  that 
these  clumps are tracing YSOs with protostellar outflows. 
Thus,  our observations  reveal 
earliest  stages  of massive star formation near Sgr A* on a time scale of $\sim10^4-10^5$ years. 
Additional support for 
star formation on a  time scales of $\sim10^5$ 
years come from  SED fitted YSO candidates with infrared excesses on $\sim10^6$ year timescales, 
as well as young stellar disks orbiting Sgr A*, respectively.  These suggest that 
star formation is continuous near Sgr A*.

The mechanism by  which star formation can 
take place in this tidally stressed environment is not well understood. 
The H$_2$ density in the molecular ring, $\sim 10^6-10^7 \rm cm^{-3}$, is well 
below that needed for self-gravity to overcome the tidal field of the 
central black hole, i.e.  $2\times 10^8 (\rm r/1 pc)^{-3}$ cm$^{-3}$.  
Star formation in this region  must therefore be triggered by 
significant compression of the ambient gas in the ring. 
Tidal squeezing  of an elongated infalling cloud can compress the gas in two dimensions, but 
the density needs to be  
 increased  by two orders of magnitude as the cloud approaches Sgr A*.  
Here we consider two other   possibilities: (i) compression by the intense UV 
radiation field in the Galactic center, and (ii) clump-clump collisions. 

Hot stars in the central parsec produce an intense radiation field with an effective temperature $\sim3\times10^4$K and a 
luminosity of order $\sim10^{7.5}$ L\solar (Lacy et al. 1980; Davidson et al. 1992). The radiation pressure is capable of 
producing significant compression.  Equating the radiation pressure to the thermal pressure of the compressed molecular gas, 
i.e. $\rm L/(4\pi\, c\, D^2) = 1.2 \rm n_{H_2} \rm k \rm T$, where D is the distance to the center of the hot star 
distribution, yields n$_{H_2}\sim4\times10^8 (\rm L/10^{7.5} L\solar) (\rm D/0.1 pc)^{-2} (\rm T/100K)^{-1}\, \rm cm^{-3}$, 
comparable to the critical density.  It is therefore possible for the pressure associated with irradiation by the UV field to 
compress a gas clump to the point of gravitational collapse.  Note, however, that the clump must be exposed to the radiation 
field for about 5$\times10^4$ years for the compression to work its way through the entire clump and that much of the momentum 
may be deposited in a photoevaporative outflow.

Clump-clump collisions are an alternative method for producing SiO emission, either through the destruction of dust grains in large 
scale shock waves associated with the collision, or via outflows from YSOs formed, 
 because the compression associated with the shock 
waves triggered star formation.  This model, however has difficulty in producing the number of detected sources.  
Suppose a clump in the  interior to the ring has  
radius r, clump-clump velocity dispersion v and volume filling factor f.  
Then the number of 
clumps per unit volume is n$_{cl} = f/(4/3 \pi r^3)$, the cross section for almost head-on collisions is $\sigma \sim 2 \pi r^2$, 
and the time 
scale for a given clump to collide with another is $1/(n_{cl} \sigma v)$.  
For an ensemble of N clumps, the number of collisions per unit 
time is  $N\,n_{cl} \sigma v \approx 1.5\times N\, f\, v / r$.  
The collisional interaction time is approximately $2r/v$ , so the number of collisions 
occurring at any given time is $(N\, f\, v\, / r) \times (2r/v) = 3f\, N$.  
For the interior of the molecular ring with a  radius of 15$''$, a clump size 1.5$''$, 
$f=10^3/N$ and the number of visible clumps is 
$3Nf=11$, thus a  population of 60 clumps are needed to produce the number of observed SiO clumps.

In conclusion, ALMA observations show that 
the interior of the circumnuclear molecular ring is not 
completely filled with  ionized gas (see the review by Genzel et al. 2010)
but   is a site of on-going  star formation. 
The linewidths of these  clumps are too large to be 
gravitationally bound, thus suggesting outflows from 
YSOs.  
The SiO clumps  we found in the Galactic center are 
highly excited  but show properties that are similar to 
to those  found in star
formation regions.  
We suggest that 
the required high gas density  is produced  by 
the strong external radiation field from young massive stars 
compressing the gas, thus inducing  star formation or that 
clump collisions can account for  compressing the gas.  
Future observations 
will determine  
the total mass of molecular gas residing inside the  ring, 
will allow estimating the efficiency of star formation 
within a pc of  Sgr A* and will examine if 
the molecular gas inside the ring is  dynamically 
important in  perturbing the dynamics of stars
close to Sgr A*.

We thank Stefan Gillessen for providing us with VLT images. 
This paper makes use of the following ALMA data: ADS/JAO.ALMA\#2011.0.00005.SVProject 
code. ALMA is a partnership of ESO (representing its member states), NSF (USA) and NINS 
(Japan), together with NRC (Canada) and NSC and ASIAA (Taiwan), in cooperation with the 
Republic of Chile. The Joint ALMA Observatory is operated by ESO, AUI/NRAO and NAOJ.

\newcommand\refitem{\bibitem[]{}}

\begin{figure}
\center
\includegraphics[scale=0.35,angle=0]{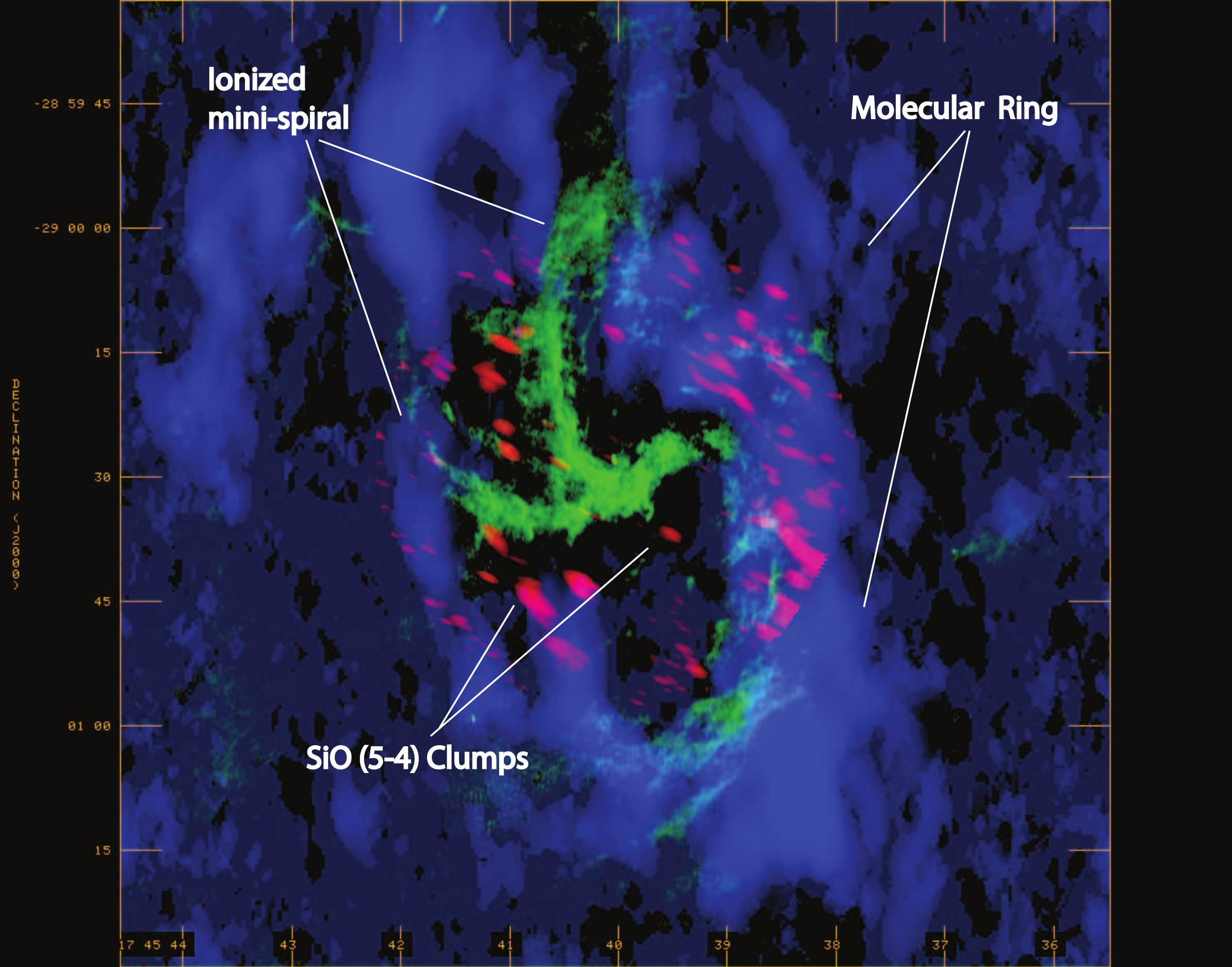}\\
\includegraphics[scale=0.35,angle=0]{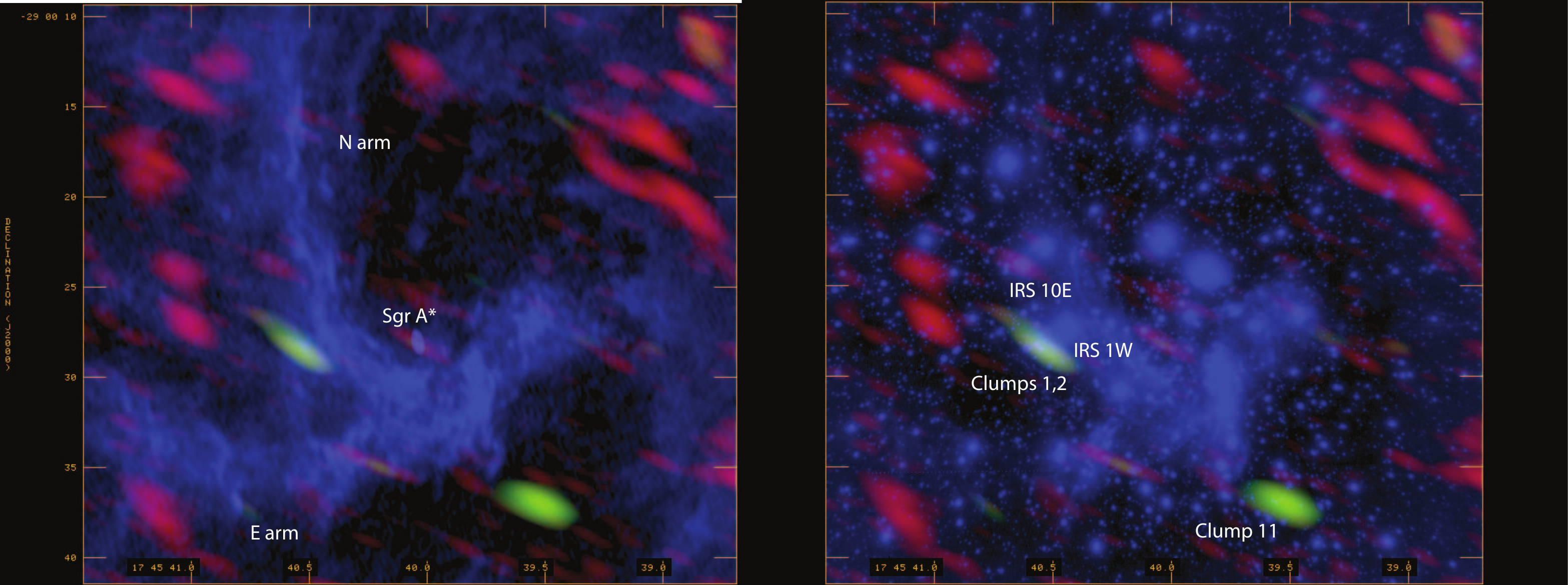}
\includegraphics[scale=0.75,angle=0]{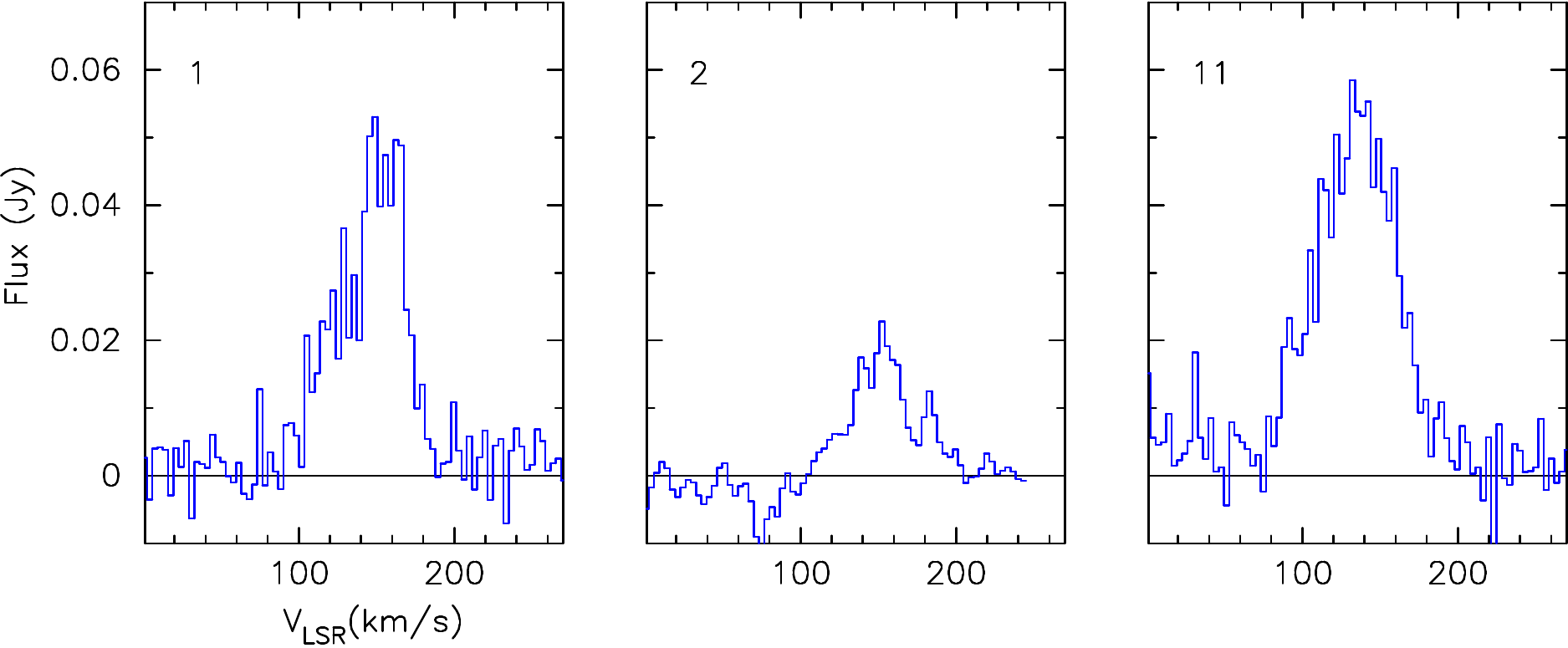}
\caption{
{\it (a)} 
A  composite image of the inner 2$'$ of the Galactic center: 3.6 cm continuum image of Sgr A West (green), 
HCN (1-0) emission from the molecular ring (blue; Christopher \etal 2005) 
and SiO (5-4) line emission from the central region of the ring (red)   
{\it (b) } 
The distribution of SiO line emission  (red) 
integrated over velocities $150 < \rm v <200$ \kms\, and superimposed on a 
3.6 cm 
continuum image (blue). 
The edge of the SiO (5-4) clump distribution is noisier because of shorter integration time and is 
limited by the size of the region mapped by ALMA. The highest redshifted velocity 
SiO (5-4) clumps 1 and 11 are shown in green. 
{\it (c)} 
Similar to (b) except that 
a 3.6$\mu$m image taken with the VLT in blue replaces the 3.6 cm image.  
{\it (d)} 
The spectra of Clumps 1, 2 and 11.}
\end{figure}  


\begin{figure}
\center
 \includegraphics[scale=0.9,angle=0]{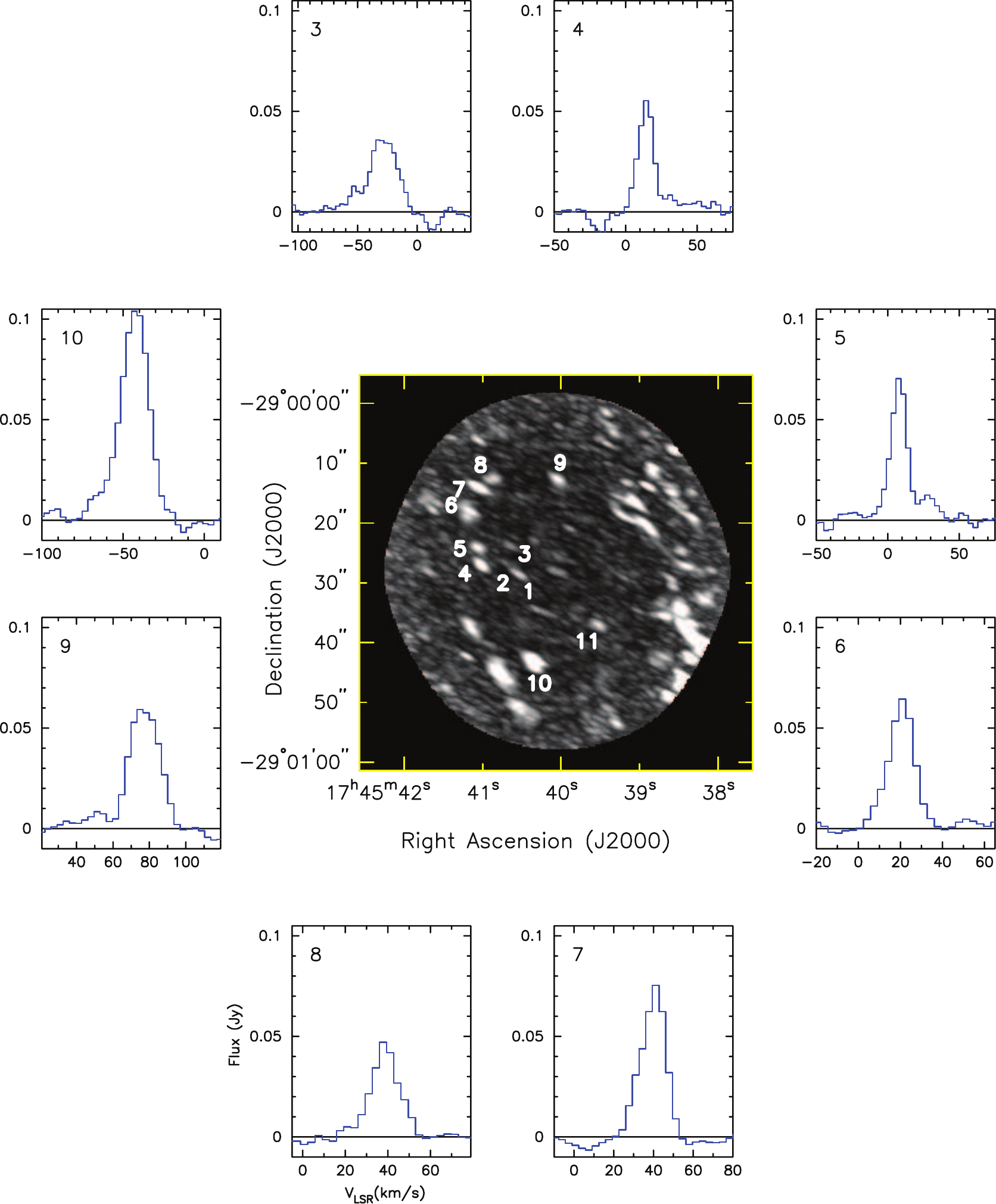}
\caption{
A grayscale image 
constructed from the peak SiO (5-4) line emission 
between $-191 < \rm v <213$ \kms. 
The inset shows the spectra of 8 SiO (5-4) clumps  (Jy on Y-axis vs \kms\, on the X-axis) within the 
molecular ring.  The positions of labeled  spectra are listed in Table 1. 
}\end{figure}


\begin{figure}
\center
 \includegraphics[scale=0.3,angle=0]{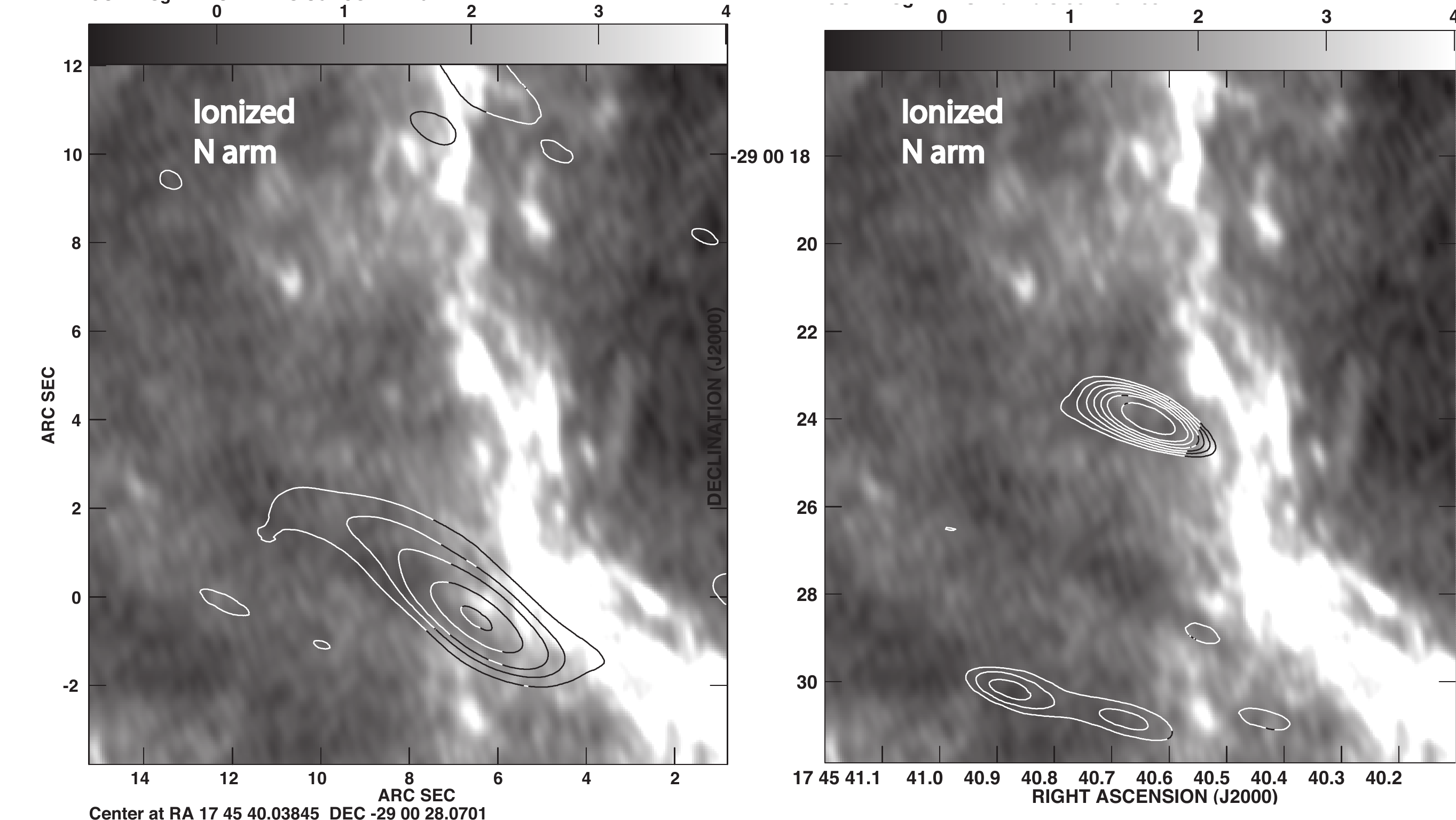}\\
 \includegraphics[scale=0.3,angle=0]{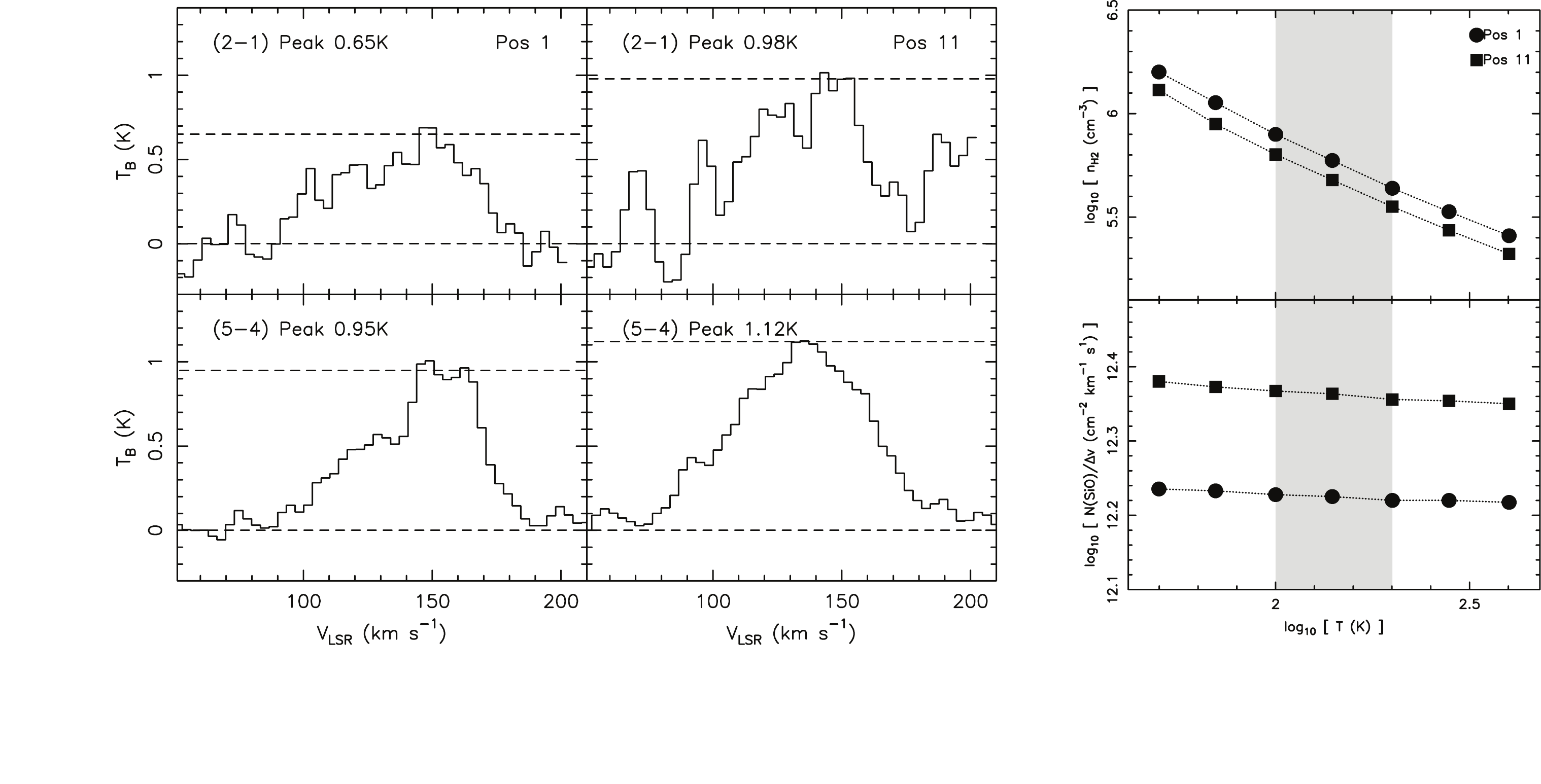}\\
 \includegraphics[scale=0.3,angle=0]{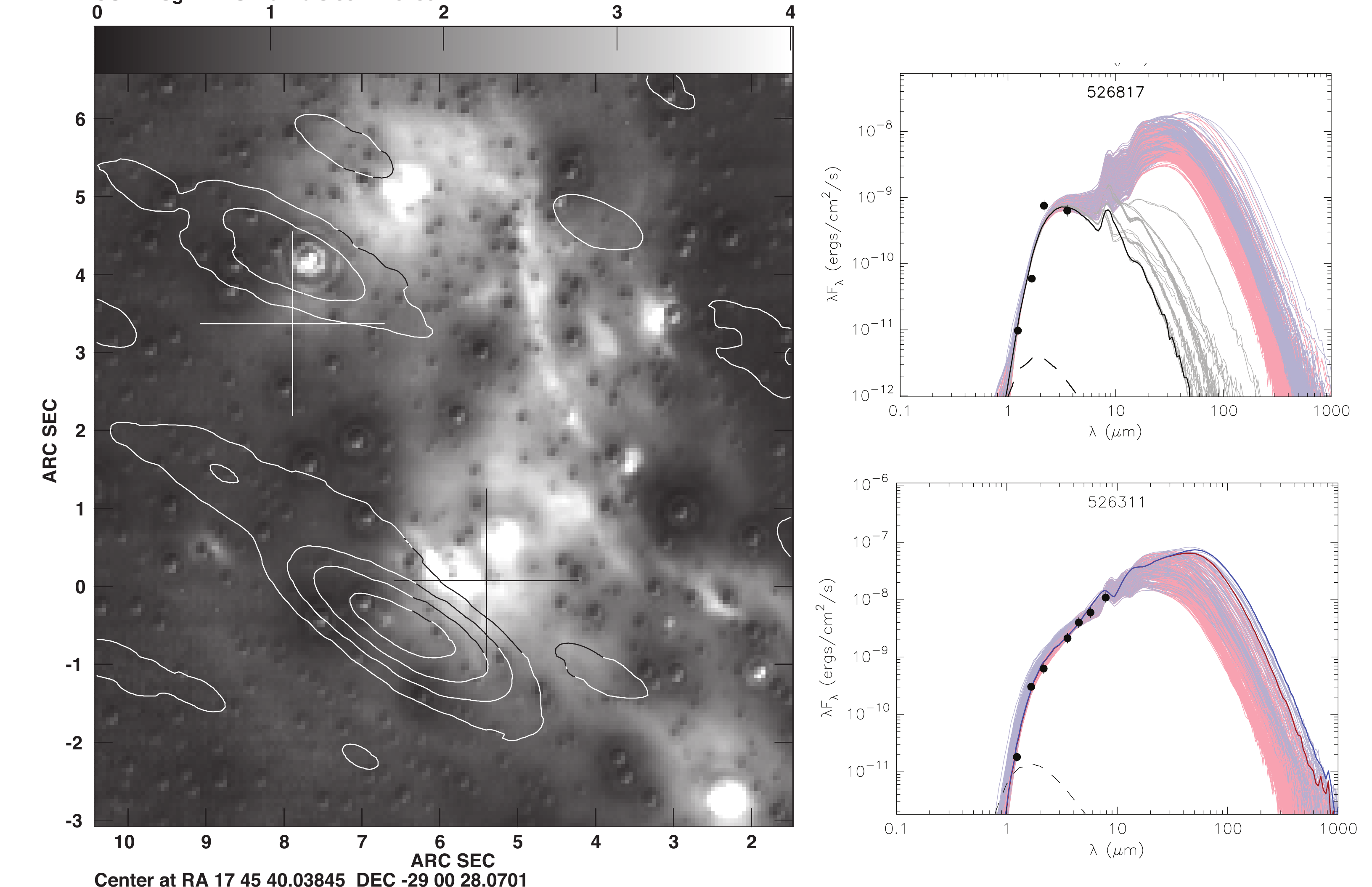}\\
\caption{
(a) Contours of SiO (5-4)  line 
emission  from Clumps 1 and 2 integrated between 88 and 188 \kms\,
superimposed on a 3cm continuum image. 
(b) Same as (a) except contours of SiO (5-4) line emission
integrated over 0 and -50 \kms\, 
(c) SiO (5-4) and (2-1)  line  profiles of Clumps 1 and 11 
(labeled as POS 1 and 11).  
(d) The inferred H$_2$ density (Top) 
and N(SiO)/$\Delta$v  (bottom) as a function 
of temperature. The band (gray) shows the temperature range of the CNR. 
(e) Contours of SiO(5-4) emission from Clumps 1 and 3 superimposed on 
a ratio map of L (3.6$\mu$m) to K (1.6 $\mu$m) bands. 
The crosses 
show the positions of  YSO candidates.
(f) Fitted SEDs of the two  YSO
candidates  526311 
and 526817 in the vicinity of Clumps 1 and 3, respectively.  
526311 and 526817 are designated in the catalog by Ramirez et al. 2008.
} 
\end{figure}  


\begin{figure}
\center
 \includegraphics[scale=0.9,angle=0]{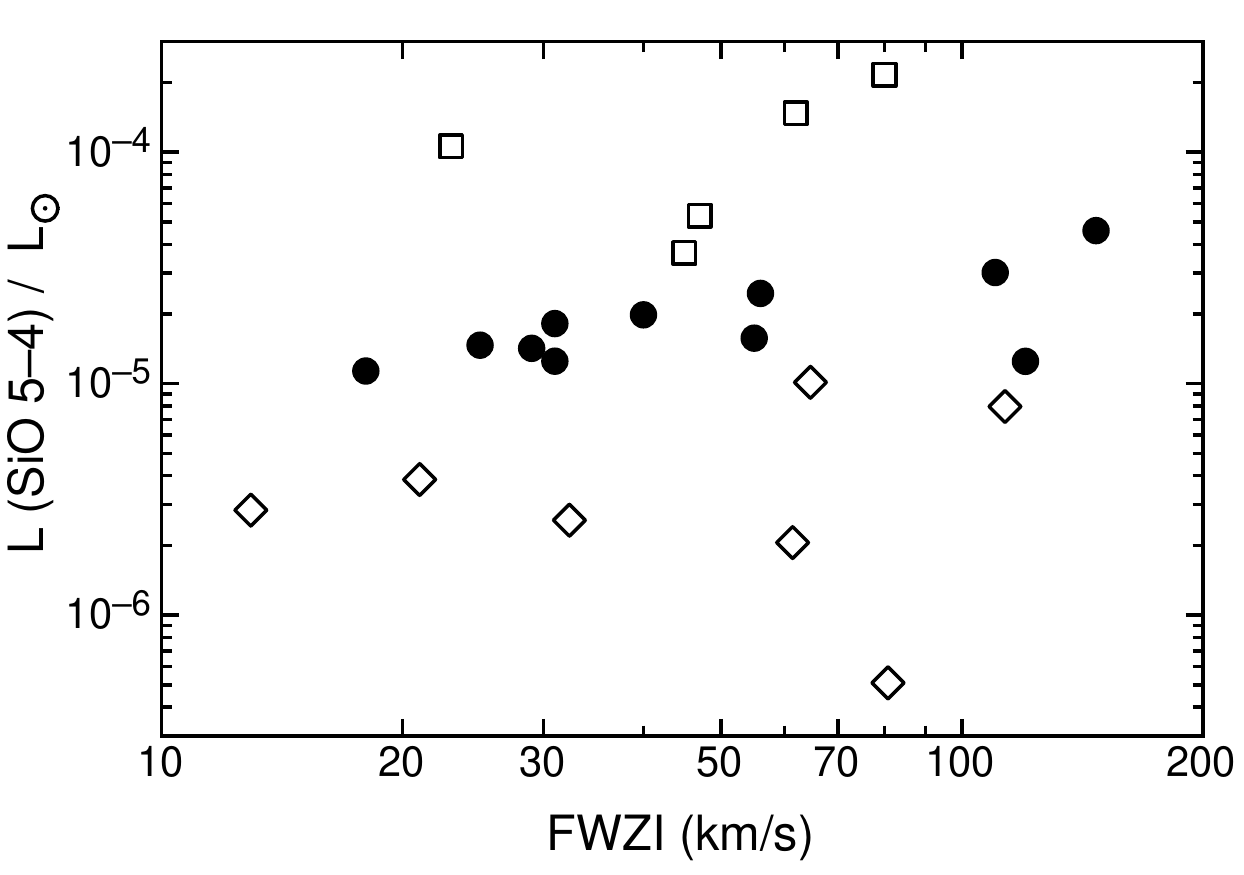}
\caption{
Filled circles show the luminosity in the SiO(5-4) line versus full line width (FWZI) 
for the 11 sources detected in the circumnuclear ring.  Open 
 diamonds and open squares   show the corresponding quantities for outflows 
from low-mass YSOs (Gibb \etal 2004) and high-mass YSOs (Gibb \etal 2007), respectively.
}
\end{figure}

\end{document}